\documentstyle[multicol,aps,epsf]{revtex}           
\begin{document}           
\title{Statistical ensemble of scale-free random graphs}           
\author{Z. Burda $^{1}$, J.~D. Correia $^{2}$, A. Krzywicki $^{2}$}           
\address{$^{1}$ Fakult\"at f\"ur Physik, Universit\"at            
Bielefeld, Postfach 100131,           
D-33501~Bielefeld, Germany\\           
and Institute of Physics, Jagellonian University,           
ul. Reymonta 4, 30-059~Krak\'ow, Poland}           
\address{$^{2}$ Laboratoire de Physique Th\'eorique, B\^atiment 210,            
Universit\'e Paris-Sud,           
91405~Orsay, France}            
\maketitle           
           
\begin{abstract}           
A thorough discussion of the statistical ensemble of scale-free     
connected random tree graphs is presented. Methods borrowed     
from field theory are used to define the ensemble and to study     
analytically its properties. The ensemble is characterized by      
two global parameters, the fractal and the spectral dimensions,      
which are explicitly calculated. It is discussed in detail how      
the geometry of the graphs varies when the weights of the nodes      
are modified. The stability of the scale-free regime is also     
considered: when it breaks down, either a scale is spontaneously     
generated or else, a "singular" node appears and the graphs become      
crumpled.  A new computer algorithm to generate these random graphs     
is proposed. Possible generalizations are also discussed. In      
particular, more general ensembles are defined along the same lines     
and the computer algorithm is extended to arbitrary (degenerate)     
scale-free random graphs.\par     
\end{abstract}     
     
\begin{multicols}{2}     
\section{INTRODUCTION}     
\par     
Random graphs are entities one encounters in many            
fields of research. Every time one has some objects           
or agents in mutual interaction, one can consider these           
objects or agents as the nodes (vertices) of a graph and represent           
pictorially the existence of the interaction by an abstract            
link (edge) connecting two nodes. Hence, an epidemic can            
be regarded as a graph: the nodes are the infected people            
and the links connect those who have been infected with            
their infectors. Likewise, the science citation index            
can be represented by a graph. Of course,           
the world-wide-web is a graph. There are many more such           
examples. Usually the graph structure is fairly random.           
\par           
Consider a random graph. Let $n$ denote the degree           
of a node and $P_n$ the corresponding bulk probability           
distribution. When $P_n \sim n^{-\beta}$ for $n \gg 1$           
it is common to call the graphs {\em scale-free}.           
This class of graphs is not described by           
the classical theory of random           
graphs \cite{erdos}, where $P_n$ is Poissonian.            
The subject is not really new, but it became popular           
recently, when new data on large graphs - in particular on            
the web network structure \cite{data} - became available.           
\par           
Scale-free graphs are naturally generated by stochastic           
processes, where the graph size is growing and the addition            
of links is preferential: the probability to attach a new link            
to a node is, roughly speaking, proportional to the node      
degree. Several models of growing           
networks have been proposed \cite{bar1}-\cite{bornh}.           
The prototype is the model worked out in a classic paper by           
Simon \cite{simon}. This model does not really deal with graphs,           
but can easily be converted into a (directed) graph model, as           
observed in \cite{bornh}.           
\par           
In studying random graphs one can adopt two complementary           
approaches:           
\par           
- The {\em diachronic} approach, where one focuses on the time            
evolution of the graph. The main advantage of            
this approach is that one            
stays close, in spirit at least, to the            
dynamics operating in Nature.           
Furthermore, one can discuss processes, like            
aging, which are intimately           
related to the time evolution. To our knowledge this            
approach is the one            
commonly used in works on scale-free graphs.           
\par           
- The {\em synchronic} approach, where one considers the statistical            
ensemble of graphs at a fixed large time. Introducing a           
statistical ensemble enables one to use the conceptual           
tools of statistical mechanics. The dynamics of processes producing           
scale-free networks is presumably much more complicated            
than that of the proposed simple models. The           
resulting randomness is better implemented in the synchronic           
than in the diachronic approach. Generating graphs with an a            
priori given connectivity distribution seems also easier.            
This is the approach adopted in this work. 

\begin{figure}     
  \narrowtext \hskip 0.6in \epsfxsize=2.3in     
\epsfbox{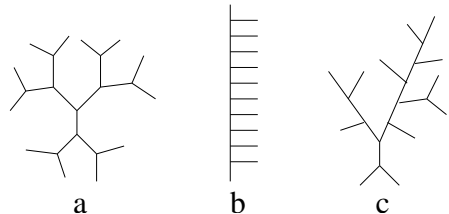} \vskip 0.1in     
\caption{Three graphs with the same number of identical nodes and     
therefore, in our model, with the same weight. A Cayley tree (a) has     
always a diameter $\sim \log{N}$, where $N=$ \#nodes. A comb-like tree     
(b) has a diameter $\sim N$. The generic trees, like (c), are devoid     
of any symmetry and can be drawn in a variety of manners, ie have     
large entropy. Such trees made up with nodes of degree one and      
three have the Hausdorff dimension $d_H=2$ (cf [12]) and      
therefore their diameter is typically $\sim \sqrt{N}$. }     
\end{figure}           
\par           
Introducing a statistical ensemble of      
graphs means ipso facto that           
the graphs are weighted. In our      
model each graph is given a weight      
proportional to the product of weights      
attached to individual nodes and depending     
on node degrees. The latter weights are      
chosen so as to generate the 
scale-free behavior in the ensemble.     
Thus, in a sense, the graphs - the microstates of the      
ensemble - are given an "internal energy".      
Our general results characterize, of course,           
the "typical" graphs only, ie those            
contributing most significantly to           
the ensemble averages. 
This can be considered as a virtue or as           
a flaw of the approach, depending on the goal one is pursuing.           
The "typicality" of graphs
results from the usual interplay of energy and
entropy (see also Fig. 1).
\par           
A rather general graph ensemble is defined in      
Sec. II. For definiteness, we consider      
undirected graphs only. We use techniques borrowed           
from field theory, which enable one to formulate the problem and to           
find a number of results in a rather elegant fashion.     
The ensemble is defined for a given set of input weights.     
      We rapidly focus on     
scale-free {\em connected tree} graphs,      
ie connected graphs without cycles. This is      
the case we have fully under control and where many      
results can be obtained      
analytically, so that a maximum clarity is achieved     
and our approach is exposed the best.     
We discuss the condition that guarantees that an a priori given            
connectivity distribution is found in the "output", ie when the           
weights of graphs and their entropy are taken            
into account. In Sec. III we show that the ensemble can            
be in a variety of phases. For each of these           
phases we calculate two important scaling exponents: the fractal and the           
spectral dimensions. In Sec. IV we discuss what happens when the input           
weights are modified. We show that the           
scale-free regime is unstable. Generically, the ensemble develops two           
responses: either a scale is spontaneously generated or else a "singular"            
node appears, which is connected to almost all other nodes (the            
"one-takes-it-all" scenario) \cite{foot2}.            
\par     
A graph model can be used as an event      
generator in Monte Carlo studies of      
the statistical properties of graphs themselves      
or of the "matter" (eg Ising      
spins, complex spins, etc) living on the      
nodes  \cite{foot1}. Our     
discussion will also lead to the formulation of      
a simple computer algorithm for generating      
random graphs. Readers      
more interested in this aspect of      
this work than in the more formal      
discussion to follow can jump directly to Sec. V, where     
our algorithm is formulated, after reading the beginning      
of Sec. II to get acquainted with our notation. In Sec. VI we present      
a sample of curves representing     
the connectivity distributions calculated      
for finite systems using our code,     
compared to the analogous curves generated      
from the growing network recipe      
of refs. \cite{bar1,doro0,krap1}     
and we briefly discuss some finite size effects.       
\par     
In Sec. VII we summarize            
our results and we discuss the            
problems related to the generalization      
of the present approach to           
models more complicated than the     
connected tree model discussed      
at length in the previous sections.      
We also list some open questions.     
In this paper, when necessary, we make      
use of some results, found in a different           
context, scattered in earlier publications we have co-authored           
\cite{bb1,bb,jk,cw}. We believe, that it is useful to adapt these           
results to the present context, putting them in a new perspective           
and making them accessible to a different community.            
           
\section{MINIFIELD THEORY AND GRAPH ENSEMBLES}           
           
Our starting idea is to use a toy field theory            
in 0 dimensions - we call           
it a minifield theory - and to identify the Feynman            
diagrams of this theory            
with the graphs of our ensemble. The weights            
attached to the graphs are            
the amplitudes calculated using the      
standard Feynman rules, the model            
being, of course, formulated in such a      
manner that these weights are       positive.           
\par           
Let us define the minifield theory \cite{jk}      
by the following formal       integral           
\begin{equation}           
Z   = (2\pi \kappa \lambda)^{-\frac{1}{2}}     
\int d\phi\; \exp{\frac{1}{\kappa} [- \phi^2/2\lambda +      
\sum_{n>0} p_n \phi^n]}           
\label{start}           
\end{equation}           
      By assumption the real constants $p_n$ are      
non-negative. Hence, strictly speaking,      
the integral does not exist. However,      
expanding the second term in the exponential     
and performing the integrations over $\phi$      
one obtains an unambiguously defined     
series. The terms of this series can be      
represented, as usual in field theory,     
by Feynman diagrams. The convergence of      
the series is, in general, uncertain and requires      
some consideration. When the      
series converges, its sum, also denoted by $Z$,      
can be regarded as the partition     
function of a graph ensemble. 
     
\begin{figure}     
  \narrowtext \hskip 0.6in \epsfxsize=2.3in     
\epsfbox{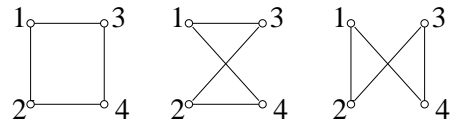} \vskip 0.1in     
\caption{The labeled connected Feynman diagrams     
representing the contributions to the rhs of (\ref{ex1}).     
With each one of these graphs is associated the weight     
$2[p_2 \lambda]^4/3$. }     
\end{figure}
\par     
The content of the above paragraph     
will appear obvious to the practitioners of     
field theory, but may appear     
fortuitous to some readers working on      
networks, but not totally familiar with      
the former. We have no space to explain     
the point at length. Hopefully, a simple     
example may be of some help. Consider     
the following contribution to $Z$:      
\begin{equation}     
Z_{\mbox{\rm \footnotesize example}} =     
(2\pi \kappa \lambda)^{-\frac{1}{2}}      
\int d\phi\; e^{- \frac{1}{2 \kappa \lambda} \phi^2}      
\; \; \frac{1}{4!} \; [p_2 \phi^2/\kappa]^4     
\label{ex1}     
\end{equation}     
We need to calculate the eighth moment of     
a Gaussian. We shall do it in a fancy way.     
For book-keeping purposes we distinguish     
by a label the four "interaction terms" that     
are being multiplied above:     
\begin{equation}     
[\phi^2]^4 \rightarrow      
\phi(1)^2 \phi(2)^2 \phi(3)^2 \phi(4)^2     
\end{equation}     
and we represent each label by a point in a     
plane. We observe that every moment of      
a Gaussian is fully determined by its      
second moment, which in our case equals      
$\kappa \lambda$. We replace all binomials     
appearing in the integrand,      
$\phi(i) \phi(j) \rightarrow \kappa \lambda$ and we     
join with a line the points representing     
the labels $i$ and $j$. The latter results in a     
specific graph, a Feynman diagram. The moment to     
be calculated is obtained by summing over     
all possible matchings of four pairs of labels.     
All the {\em connected} graphs corresponding      
to our example are shown in Fig. 2. One     
checks easily that to each of these diagrams     
correspond 16 distinct matchings of the same four pairs     
of labels. Thus, the contribution to the rhs of      
(\ref{ex1}) of each of these Feynman diagrams is      
$\lambda^4 [p_2 2!]^4/4!$ . The full     
result is obtained by taking into account the     
disconnected diagrams too. In some diagrams     
a line connects a point to itself ($i=j$),     
these are the so-called "tadpoles". We     
shall not list all these diagrams here.     
\par     
The point is that the Feynman diagrams of the      
minifield theory are the graphs familiar to     
people working on networks, except that there is a     
specific weight attached to each such a graph.     
The graphs are not necessarily connected, except     
if one imposes an appropriate extra constraint,     
defining a graph sub-ensemble.     
\par     
We assume that $p_1$,      
which plays the role of an     
external current, is strictly positive. Hence,      
in our minifield theory there exist graphs with     
nodes that are "external", ie of unit degree.     
\par           
The weight associated with a Feynman diagram            
that is  "non-degenerate" -            
ie does not have tadpoles and multiple            
connections between nodes -  and            
whose nodes are labeled, is            
\begin{equation}           
\mbox{\rm weight} = \kappa^{L-N} \; \frac{\lambda^L}{N!}\;            
\prod_{j=1}^N \; [\; p_{n_j} n_j! ]           
\label{w}           
\end{equation}           
\par\noindent           
where $N$ and $L$ denote the total number      
of nodes and links respectively.            
When tadpoles and/or multiple            
connections are present one has to multiply the            
rhs by the usual symmetry factors.            
In graph theory some authors accept      
the existence of degeneracies in the            
definition of what they mean by a graph. In most            
texts the degeneracies are            
excluded. The class of graph models we      
propose is fairly general, but not     
the most general. Indeed, by construction     
the weight of a graph is given in      
(\ref{w}) by a product of weights     
associated with individual nodes. A possible     
generalization will be proposed later on.      
\par           
As is well known from field theory, $W=\kappa \ln{Z}$     
generates the connected graphs. Only tree graphs survive     
in the "semi-classical" limit            
$\kappa \to 0$. From here on we focus on the     
statistical ensemble of {\em connected tree      
graphs}. We shall return to     
more general graphs in the last section.     
\par     
In this paper we are interested in models where      
the radius of convergence of      
the series in the exponent on the rhs of      
(\ref{start}) is finite. It is not      
a loss of generality to set this radius to unity. It is evident      
from (\ref{w}) that this can always be achieved by a multiplicative      
renormalization of $\lambda$ and $\kappa$.     
\par     
The partition function of connected tree graphs     
$W^{\mbox{\rm \footnotesize trees}}$ is given            
by the stationary value of the            
action at $\phi=\Phi$, $\Phi$ being defined            
implicitly by the saddle-point            
equation           
\begin{equation}           
\Phi = \lambda \sum_{n>0} t_n \Phi^{n-1}           
\label{root}           
\end{equation}           
\par\noindent           
where we have introduced a shorthand notation $t_n=np_n$.           
One easily checks that           
\begin{equation}           
\Phi = \partial W^{\mbox{\rm \footnotesize trees}}/\partial t_1             
\end{equation}           
\par\noindent           
Hence $\Phi$ generates tree graphs with one marked external node. Eq.            
(\ref{root}) with its nice graphical            
representation \cite{jan,nsw} - see Fig. 3            
 - has a long history and is recurrently rediscovered in the            
literature. 
     
\begin{figure}     
  \narrowtext \hskip 0.6in \epsfxsize=2.3in     
\epsfbox{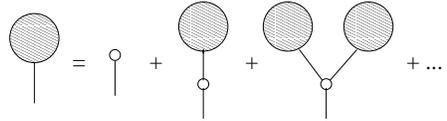} \vskip 0.1in     
\caption{The graphical representation of eq. (\ref{root}).     
With every node in the figure - the open circles - is associated the     
factor $\lambda t_n$. We use the shorthand notation $t_n=np_n$.  }     
\end{figure}           
\par           
It is easy to see that $\Phi$ is singular. Indeed, eq. (\ref{root}) can            
only be satisfied if $\lambda$ is smaller than some critical value            
$\lambda_c = \lambda_c(t_1, t_2, ...)$.            
Equation (\ref{root}) can be rewritten as           
\begin{equation}           
\lambda^{-1} = F(\Phi) \equiv \sum_{n>0} t_n \Phi^{n-2}           
\label{saddle}           
\end{equation}           
\par\noindent           
and by definition           
\begin{equation}           
\lambda_c^{-1} = F(\Phi_0)            
\label{sad}           
\end{equation}           
\par\noindent           
where $\Phi=\Phi_0$ is the point in the            
interval $[0,1]$ where $F(\Phi)$           
takes its minimal value. The critical behavior of the theory is found           
from the behavior of $F(\Phi)$ in the vicinity of $\Phi_0$.            
\par           
One observes that both the rhs of (\ref{saddle})           
and its second derivative are positive            
for $\Phi > 0$. Moreover, $\lambda_c$ is finite (because $t_1 > 0$).           
The concavity property of $F(\Phi)$           
implies that one has, in general, the choice between only           
three possibilities (cf refs. \cite{bb1,jk,jan})~:           
\vspace{0.3cm}           
\par           
(A) $\Phi_0 < 1$ and $F'(\Phi_0) = 0$.           
\par           
(B)  $\Phi_0 = 1$ and $F'(1) = 0$.           
\par           
(C)  $\Phi_0 = 1$ and $F'(1) < 0$.           
\vspace{0.3cm}           
\par\noindent           
These three cases will be discussed in detail later on.            
\par           
It is customary to introduce a susceptibility critical exponent           
$\gamma$ controlling the most singular part of the second derivative of           
the partition function. Since           
$\Phi$ is the first derivative of the partition function, we write           
\begin{equation}           
\Phi \sim (\lambda_c - \lambda)^{1-\gamma}           
\end{equation}           
\par\noindent           
The exponent $\gamma$           
is finite in all cases of interest for us, as will be shown in the            
following sections, where it will be calculated.           
\par           
A factor $\lambda$ is associated with every link of a tree graph and            
has the physical significance of the fugacity of links. Since           
$L=N-1$, it can also be regarded as the fugacity of nodes. Hence           
\begin{equation}           
W^{\mbox{\rm \footnotesize trees}} =            
\sum_N W_N^{\mbox{\rm \footnotesize trees}} \; \lambda^N            
\sim (\lambda_c - \lambda)^{2-\gamma}           
\end{equation}           
\par\noindent           
Taking the inverse Laplace transform one obtains for tree graphs            
the canonical partition function           
\begin{equation}           
W_N^{\mbox{\rm \footnotesize trees}}           
 \sim N^{\gamma-3} \lambda_c^{-N}  \; \; \; \; N \gg 1           
\label{entr}           
\end{equation}           
\par\noindent           
Hence, the           
ensemble of random trees behaves like a            
standard statistical mechanical           
ensemble. In particular, the number of            
trees with $N$ nodes is exponentially            
bounded in $N$ and the canonical ensemble has an extensive entropy.           
\par           
We have not yet introduced the constraints associated with the           
absence of a scale. The quantity of experimental            
interest is $P_n$, the bulk            
connectivity distribution. It is $P_n$ which should exhibit a power-like           
fall at large $n$. The relevant question            
is how to choose the series of the             
bare couplings $p_n$ in order to get a            
given scale-free $P_n$. The rest of this section is devoted            
to this problem. For tree graphs           
the solution can be found mapping the model of            
graphs on another one, known in the           
literature as the balls-in-boxes model \cite{bb}.           
\par           
In Sec. 2.7.7 of ref. \cite{book} a series solution of eq.            
(\ref{root}) is given :           
\begin{equation}           
\Phi = \sum_N \Phi_N \; \lambda^N           
\end{equation}           
\par\noindent           
with           
\begin{equation}           
\Phi_N = \frac{1}{N} \sum_{\{v_j\}} \frac{N!}{v_1! v_2! ... v_N!} \;           
t_1^{v_1}\; ... \; t_N^{v_N}           
\label{root2}           
\end{equation}           
\par\noindent           
and $v_j$ being positive integers satisfying the constraints           
\begin{equation}           
\sum_j v_j = N    \; \; \; \; \; \sum_j j v_j = 2N - 1           
\end{equation}           
\par\noindent           
Eq. (\ref{root2}) can be further rewritten as           
\begin{equation}           
N \Phi_N = \Omega(N, 2N-1)            
\end{equation}           
\par\noindent           
with           
\begin{equation}           
\Omega(N, M)  = \sum_{\{n_j\}} t_{n_1}...t_{n_N} \; \delta            
(\sum_{j=1}^N n_j - M)           
\label{binb}            
\end{equation}           
\par\noindent           
The summation is now on all sets of $N$ positive integers. The rhs            
is the partition function of a system of $M$            
balls distributed at random            
and with weights $t_n$ among $N$ boxes, empty boxes being forbidden.            
This model of weighted partitions is precisely            
the balls-in-boxes model\cite{foot3}.           
\par           
The effective box occupation distribution is           
$t_n \Omega(N - 1, M - n)/\Omega(N, M)$.            
Using the integral representation of            
the (discrete) $\delta$-function one           
easily finds an integral representation            
of $\Omega(N,M)$, which can be used           
to calculate it by the saddle-point            
method in the limit $N \to \infty$,           
$\rho = M/N = \mbox{\rm const}$. The            
effective box occupation distribution            
follows. Of course, the tree graph model            
corresponds to $\rho = 2$. We shall            
not repeat this calculation here, referring the reader            
to ref. \cite{bb} for details and quoting the result for            
$\rho = 2 \leq \langle n \rangle_t \equiv \sum_n n t_n/\sum_n t_n$ :           
\begin{equation}           
P_n \sim t_n \Phi_0^n           
\label{int}           
\end{equation}           
\par\noindent           
up to an obvious normalization factor. Here $\Phi_0$            
is to be found from            
$F'(\Phi_0) = 0$. This last condition can be rewritten as           
\begin{equation}           
\sum_n n t_n \Phi_0^n/\sum_n t_n \Phi_0^n = 2           
\end{equation}           
\par\noindent           
The lhs is an increasing function of $\Phi_0$.            
Consequently $\langle n \rangle_t = 2$           
corresponds to $\Phi_0=1$ and therefore to the case (B)           
introduced earlier. Hence, for tree graphs and in the limit            
$N \to \infty$            
\begin{equation}           
P_n \sim t_n   \; \; \; \mbox{\rm iff} \; \; \;            
\langle n \rangle_t = 2           
\label{cond}           
\end{equation}           
\par\noindent           
This result can be understood intuitively. The shape of the two           
distributions, $P_n$ and $t_n$, differs because of the constraint            
$s_N \equiv \frac{1}{N} \sum_j n_j = 2$. But for $N \to \infty$            
the constraint is satisfied "for free" when $\langle n \rangle_t =2$            
and $\beta > 2$ by virtue of Khintchin's law of large numbers.      
      \par           
The condition (\ref{cond}) solves our problem:            
the input couplings $p_n$ of the            
tree-graph model should be set to $p_n = P_n/n$            
in order to get the scale-free           
connectivity distribution $P_n$ in            
the bulk. The distribution $P_n$           
must, of course, satisfy the constraint $\sum_n n P_n = 2$,            
otherwise it would not refer to trees \cite{foot4}.           
           
\section{GEOMETRY OF SCALE-FREE TREE GRAPHS}           
           
In this section we discuss the geometry of graphs belonging to the            
statistical ensemble of scale-free trees. Hence,            
according to the results           
of the preceding section, we assume that            
$t_n \sim n^{-\beta}$ for $n \to \infty$            
and that $\langle n \rangle_t =2 $.           
We have already stated in the preceding section that this           
situation corresponds to the case (B): $\Phi_0 = 1$ and $F'(1) = 0$.           
\par           
It follows from general arguments that $F(\Phi)$ has a branch point           
at $\Phi=1$. The order of the singularity is determined by the shape           
of the tail of $t_n$. For non-integer $\beta$ one finds (cf \cite{bb1})           
that\cite{foot5}           
\begin{eqnarray}           
F(\Phi) & = & \mbox{\rm polynomial in }(1-\Phi)           
 + (1-\Phi)^{\beta-1} \\ \nonumber           
& + & \mbox{\rm higher order terms}           
\label{ser}           
\end{eqnarray}           
\par\noindent           
The concavity of $F(\Phi)$ implies $\beta >2$.            
The linear term in the polynomial is absent.           
\par           
When $2 < \beta < 3$, inverting eq. (\ref{ser}) one finds the singular           
part of $\Phi$:           
\begin{equation}           
\Phi_{\mbox{\rm \footnotesize sing}} \sim           
 (\lambda_c - \lambda)^{1/(\beta-1)}           
\end{equation}           
\par\noindent           
so that $\gamma = (\beta-2)/(\beta-1)$.           
\par           
When $\beta > 3$ the inversion yields           
\begin{equation}           
\Phi_{\mbox{\rm \footnotesize sing}} \sim (\lambda_c - \lambda)^{1/2}           
\end{equation}           
\par\noindent           
Therefore, $\gamma =  1/2$.           
\par           
We now consider two scaling exponents which have a direct geometrical           
interpretation in terms of the average "dimension" of the ensemble graphs.           
\par           
One can define a two-point correlation            
function $C(x, \lambda)$ equal, up to            
normalization, to the average number of pairs of nodes separated by            
a distance $x$. Using a graphical representation           
analogous to that of Fig. 3 one finds \cite{jan}:           
\begin{equation}           
C(x, \lambda) \sim [\lambda           
 \frac{\partial}{\partial \Phi} (\Phi F(\Phi)]^x           
\end{equation}           
\par\noindent           
The fractal dimension $d_H$ of the tree can be defined by           
\begin{equation}           
x^{-1} \log C(x, \lambda) = - \mbox{\rm const}           
 (\lambda_c - \lambda)^{1/d_H}            
\; \; \;   \mbox{\rm for}  \; \; \; x \to \infty           
\label{c}           
\end{equation}           
\par\noindent           
where it is assumed that $\mid \lambda_c - \lambda \mid \ll 1$.            
Notice, that $(\lambda_c - \lambda)^{-1}$ is a measure of the average           
size of a graph\cite{foot6}.           
One finds finally           
\begin{equation}           
d_H = \frac{1}{\gamma}            
\end{equation}           
\label{nice}           
\par\noindent           
\par           
Another definition of dimension is obtained by considering a diffusion           
process on the graph; as is well known,            
diffusion can be mapped to a random           
walk. If we consider a walker departing from a certain point at time           
$T=0$, then the return probability to the            
start of the walk is, for small           
$T$, proportional to $T^{-d_s / 2}$ where            
$d_s$ is the spectral dimension: the           
dimension of the graph as seen by a random walker moving on the graph.           
\par           
In was shown in \cite{cw,jan2} that the spectral dimension $d_s$ is for            
$\gamma \geq 0$ given by           
\begin{equation}           
d_s = \frac{2}{1+\gamma}           
\label{ds}            
\end{equation}           
\par\noindent           
Combining (26) and (\ref{ds}) one obtains the nice relation           
\begin{equation}           
d_s = \frac{2d_H}{1+d_H}           
\label{twod}           
\end{equation}           
\par\noindent           
For $\beta > 3$ one finds             
$d_H=2$ and $d_s=4/3$. These are the generic values for a tree           
graph \cite{jan,jan2}, which would be            
found also if there were only a finite           
number of non-vanishing couplings $t_n$. The typical            
graph is a fully developed tree, with many branching branches.           
\par           
For $2 < \beta < 3$ one obtains:            
\begin{equation}           
d_H = (\beta-1)/(\beta-2)             
\label{add1}           
\end{equation}           
\begin{equation}           
d_s = 2(\beta-1)/(2\beta-3)           
\label{add2}           
\end{equation}           
\par\noindent           
As $\beta \to 2$ the trees become increasingly crumpled, they look           
like a set of linked hedgehogs. As $\beta$ approaches 2 the average     
distance between a pair of nodes starts growing very slowly with     
$N$. E.g. when $\beta=2.1$ this distance is $\sim N^{1/d_H} = N^{1/6}$.     
The increase is not logarithmic, as is sometimes claimed, but     
power-like. Of course, in practice this does not make much difference,     
when the power is small.     
\par           
The upshot of the above discussion is that            
a scale-free large-degree            
behavior $P_n \sim n^{-\beta}$, $\beta > 2$, is associated with            
a variety of distinct graph geometries. It is quite            
remarkable, that at least for the tree graphs            
of the discussed ensemble           
one can give a complete catalogue of expected geometries.           
           
\section{INSTABILITY OF SCALE-FREE REGIME}           
           
The logical next step of the discussion is to examine what happens           
when the shape of $t_n$ is so distorted that the constraint            
$ \langle n  \rangle_t = 2$ is broken. We continue assuming that            
$t_n \sim n^{-\beta}$ at large $n$.           
\par           
Consider first the case $\langle n \rangle_t < 2$.            
This condition is equivalent to           
$F'(1) < 0$. We recover the case (C) of Sec. II. Since           
the linear term in the polynomial on the rhs of (\ref{ser}) is now           
present, inverting eq. (\ref{ser}) one gets           
\begin{equation}           
\Phi_{\mbox{\rm \footnotesize sing}}           
 \sim (\lambda_c - \lambda)^{\beta- 1}           
\label{bla}           
\end{equation}           
\par\noindent           
with $\beta > 2$. An explicit calculation using (\ref{c})            
yields $d_H = \infty$.           
We also expect $d_s = 2$. A typical tree graph looks like a           
set of linked hedgehogs. Moreover, as explained below, a new feature           
shows up: a "singular" node with a fixed degree of $O(N)$, a hedgehog           
even more spiky than the others.           
\par           
To see this, it is instructive to understand how            
the constraint            
$s_N = 2$ is satisfied. Notice that if            
we omit from $s_N$ a few terms, say $n_{j_1},           
 \; ... \; n_{j_p}$, the resulting sum            
will fluctuate around $\langle n \rangle_t$.            
Hence, for $N \to \infty$           
\begin{equation}           
s_N - \frac{1}{N} (n_{j_1} + ... + n_{j_p})           
 = \langle n  \rangle_t           
\end{equation}           
\par\noindent           
But since $s_N = 2$, it is clear that some, at least, of the            
contributions to the sum on the lhs must be of order $O(N)$. Since the            
probability distribution falls like a power, it is most probable            
that there is only one such contribution. Thus, one expects the            
appearance of a characteristic "singular" node with degree            
$n = (2 - \langle n\rangle_t) N$. Other nodes have degree           
distribution $\sim t_n$:           
\begin{equation}           
P_n =  \frac{t_n}{\sum_k t_k} +            
\frac{1}{N} \delta [n - (2 - \langle n \rangle_t) N]           
\end{equation}           
\par\noindent           
The normalization factor in front of  $\delta$            
is determined by the tree            
condition $\sum_n n P_n = 2$.            
\par           
The other possible case is $\langle n \rangle_t > 2$.            
This condition is equivalent           
to $F'(1) > 0$, which by virtue of the            
concavity of $F(\Phi)$ implies that            
$\Phi_0 < 1$ . This is the case (A) of Sec. II. Hence            
$P_n$ equals $t_n$ times an exponentially            
falling factor. A scale in            
the node order distribution is spontaneously            
generated. The dramatic fluctuations of node     
degrees are absent.           
\par           
Expanding $F(\Phi)$ in the neighborhood of $\Phi = \Phi_0$, where its           
derivative is bound to vanish and inverting the resulting           
expansion in order to find the most singular part of $\Phi$ one gets           
\begin{equation}           
\Phi_{\mbox{\rm \footnotesize sing}} \sim           
 (\lambda_c - \lambda)^{\frac{1}{2}}           
\end{equation}           
\par\noindent           
Hence $\gamma= 1/2$ and one again obtains the generic           
values $d_H=2$ and $d_s=4/3$.           
\par           
The above two scenarios are the only alternatives            
to the scale-free behavior           
of the discussed statistical ensemble of tree graphs.           
\par           
One can show that the transitions between            
different phases of the model           
are continuous and that the degree of their softness varies             
when the couplings $t_n$ change. There is no phase transition between           
the generic phase (A) and the scale-free regime (B) with $\beta>3$.           
           
\section{COMPUTER ALGORITHM}           
           
In the synchronic approach the construction of graphs does not            
mimic any real physical process. The emphasis is on the flexibility            
and on the efficiency in producing a sample of graphs to work with.           
We use the following notation~: $R(n)=p_n/p_{n-1}$.           
\par           
We propose an algorithm working for a given fixed $N$ and $L$.            
It rewires links, generating Feynman diagrams of the minifield           
theory, proceeding as follows\cite{foot7}:           
\begin{description}           
   \item[(a)] Pick a random oriented link $\vec{ij}$.            
   \item[(b)] Pick a random target node $k$ .            
   \item[(c)] Apply the Metropolis test:            
rewire the link $\vec{ij}$ to $\vec{ik}$           
with probability            
\end{description}           
\begin{equation}           
\mbox{\rm Prob(rewiring)} = (n_k + 1) R(n_k + 1) / n_j R(n_j)           
\label{metro}           
\end{equation}           
when the rhs above is less than unity and with 
probability equal to unity otherwise.
\par
These Feynman diagrams are properly weighted. The            
algorithm is probing all possible connections     
within a set of distinguishable nodes. For non-degenerate           
diagrams the Metropolis condition immediately follows from            
the detailed balance equations           
together with the weight (\ref{w}) given to            
microstates. For general diagrams           
the symmetry factors, related to the presence            
of tadpoles and multiple           
connection between nodes, also come out correctly.            
This is again a consequence of the detailed balance~:           
if two nodes are connected by $m$ links, the transition to a           
state where the nodes are connected by $m-1$ links can be           
done in $m$ ways. The inverse operation has only one realization.           
When $p_n=1/n!$ the Metropolis test is always positive and           
the non-degenerate  labeled diagrams produced by the algorithm           
are equiprobable, as they should be.           
We have checked the validity of the general           
argument on a variety of examples.           
\par           
In order to produce connected tree graphs, one            
starts from a connected tree configuration, eg           
a polyline with free ends and one adds,            
before the Metropolis test, a check           
insuring that $k \neq i, j$ and $n_i = 1$.     
It is easy to convince oneself that with     
this constraint the tree is never broken into     
parts.       As explained in Sec. II, the           
scale-free connectivity distribution in the            
bulk is obtained setting            
$p_n \sim P_n/n$.                 
      \par     
Notice, that with a constraint added before the     
Metropolis test, the algorithm explores a well defined     
sub-ensemble of graphs. However, because the detailed     
balance is satisfied, this sub-ensemble is in equilibrium,     
or, more precisely fluctuates around equilibrium (see also the     
remark on that matter in the last section). Thus, within     
the sub-ensemble in question the algorithm tends to sample     
"typical" graphs. 
     
\begin{figure}     
  \narrowtext \hskip 0.6in \epsfxsize=2.3in     
\epsfbox{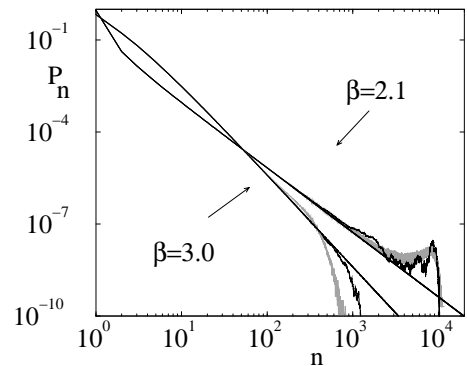} \vskip 0.1in     
\caption{Comparison of the connectivity distribution           
produced by our Monte Carlo algorithm (black) and that           
of ref. [7] (grey) for $N=2^{14}$ and $\beta=$ 2.1 and           
3.0, respectively.  The points are joined           
by a line to guide the eye. The almost           
straight lines correspond to the           
asymptotic $P_n$ given in (35).}     
\end{figure}       
     
\section{NUMERICAL EXAMPLES AND FINITE-SIZE EFFECTS}     
Since we propose a new algorithm      
for generating random graphs, it may     
be of interest to the reader to see how our results compare to those     
obtained with an algorithm already existing      
on the market. As an example     
to compare with we take the Growing Network      
model of ref. \cite{krap1},     
which is a slight generalization of the algorithm      
proposed originally in ref.     
\cite{bar1} and which also generates connected trees.      
The asymptotic analytic solution for $P_n$ is given     
by eq. (12) in the second paper of ref. \cite{krap1} (see also     
\cite{doro0}):      
\begin{equation}     
P_n = (\beta-1) \frac{\Gamma(2\beta-3) \Gamma(n+\beta-3)}     
{\Gamma(\beta-2) \Gamma(n+2\beta-3)}     
\label{kr}     
\end{equation}   
     
\begin{figure}     
  \narrowtext \hskip 0.6in \epsfxsize=2.3in     
\epsfbox{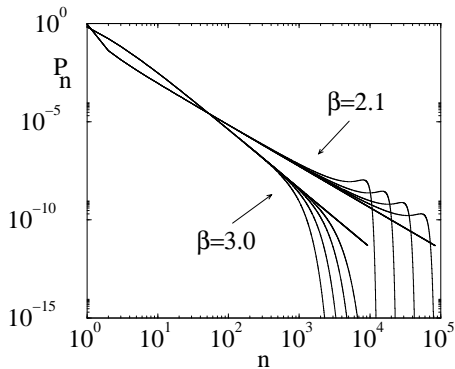} \vskip 0.1in     
\caption{The connectivity distribution calculated for           
our ensemble of tree graphs, at finite $N = 2^k , \; \;k= 14, 15,           
16, 17$ and for $\beta=2.1$ and 3.0 . The almost           
straight lines correspond to the           
asymptotic $P_n$ given in (35). The figure illustrates the           
finite-size cut-off of the order distribution discussed in the text.}     
\end{figure}    
\par\noindent     
We use this $P_n$ as input in our code. 
A sample result is shown in Fig. 4,      
where we compare the results of the calculation using the     
two algorithms, respectively.       
The calculated points      
are practically indistinguishable and     
follow the input distribution, except for those $n$ where     
finite-size effects cut and distort the distribution.     
\par     
Indeed, when $N$ is finite, the distribution of node degrees cannot     
extend to infinity, there is an upper cut-off on $n : n < n_{\mbox{\rm      
\footnotesize max}}$. The cut-off can be estimated by imposing the      
condition that the expected number of nodes found above $n_{\mbox{\rm     
\footnotesize max}}$ is at most equal to unity. Then     
\begin{equation}     
N \sum_{n_{\mbox{\rm \tiny max}}}^{\infty}     
 n^{-\beta} = \mbox{\rm const}     
\label{tail}     
\end{equation}     
\par\noindent     
The value of the constant on the rhs depends on the detailed shape     
of $P_n$. Eq. (\ref{tail}) leads to the scaling law     
\begin{equation}     
n_{\mbox{\rm \footnotesize max}} \sim N^{1/(\beta - 1)}     
\label{cut}     
\end{equation}     
\par\noindent     
analogous to the result of \cite{krap1,doro1}, where instead of     
$N$ appears the time. Beyond $n_{\mbox{\rm \footnotesize max}}$     
the node distribution function falls very abruptly and can be neglected     
for all practical purposes.      
Notice, that $n_{\mbox{\rm \footnotesize max}}$     
 may be much smaller     
than $N$. As noticed in ref. \cite{doro1} this explains why large     
scale-free graphs with $\beta$ significantly larger than 3 are not     
observed in Nature.     
\par     
In Fig. 5  we show the evolution      
of $P_n$ with $N$. In order     
to obtain a very precise result we use,      
instead of the Monte Carlo, an     
analytic recursion formula      
satisfied by the equivalent partition function     
given by eq. (\ref{binb}) of Sec. II (see footnote \cite{foot3}     
in that section).     
When the abscissa in Fig. 5 is rescaled, for      
different $N$ and using (\ref{cut}), the curves     
corresponding to the same $\beta$ are cut at the same place.       
\vspace{2cm}
     
\section{SUMMARY AND DISCUSSION}           
      \subsection{Connected tree graphs: a tentative summary}     
Let us summarize what has been achieved so far:      
We have started by introducing a fairly general ensemble     
of random graphs, but we have rapidly focused our attention      
on the sub-ensemble of connected tree graphs. We have found     
the conditions insuring that the latter are scale-free.     
We have discussed in detail the      
geometry of the connected tree graphs.     
We have also constructed a computer algorithm generating     
these graphs with an arbitrary bulk connectivity distribution $P_n$.     
Since our analytic discussion may sound somewhat            
formal for some readers, let us illustrate it with a simple, but           
very instructive example.

\begin{figure}     
  \narrowtext \hskip 0.6in \epsfxsize=2.3in     
\epsfbox{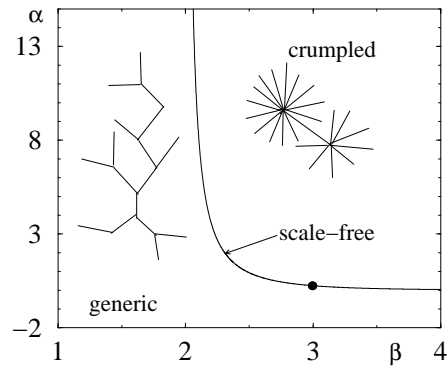} \vskip 0.1in     
\caption{Phase diagram for the simple two parameter           
model presented in Sec. VII: $t_1=1, \; \;           
 t_n = \frac{1}{\alpha} n^{-\beta} ,           
\; \; n>1$. The part of the continuous line on the right of the           
black dot corresponds to the scale-free graphs belonging to           
the generic phase. }     
\end{figure}           
\par           
Consider a two-parameter family of input weights $t_n=np_n$: $t_1=1$ and           
$t_n = \frac{1}{\alpha}\; n^{-\beta}, \; n > 1$. The condition            
insuring that the trees are scale-free was found in Sec. II            
and is $\langle n \rangle_t = \sum_n nt_n/\sum_n t_n = 2$.            
In the example in question it is equivalent to the following relation           
between the parameters $\alpha$ and $\beta$:           
\begin{equation}           
\alpha =  1 + \zeta(\beta-1) - 2\zeta(\beta)           
\label{eqab}           
\end{equation}           
\par\noindent           
where $\zeta(\beta)$ is the Riemann            
Zeta function. The phase diagram \cite{bb1}           
is shown in Fig. 6 : the line is calculated from (\ref{eqab}).           
\par           
As one moves along this line in the direction of decreasing $\beta$,           
the geometry remains generic down to $\beta=3$ and becomes progressively           
more and more crumpled below that point. In particular,            
the fractal dimension increases from the initial value           
2 to $\infty$ (see eq. (\ref{add1})).            
This behavior is           
easily understood. When $\beta$ is moved            
towards 2 the convergence of           
the series deteriorates. The only way to            
keep the average $\langle n \rangle_t$           
constant, viz. equal to 2, is to reduce the normalization of the           
tail. This means, however, that the relative weight of nodes with            
unit degree gets larger. Eventually, the number of "external" nodes           
increases with $N$ much faster than the number of large degree nodes.            
The latter are usually connected to the former, building           
hedgehog-like structures.           
\par           
Off the line (\ref{eqab}) one falls into one of the alternative           
phases discussed in Sec. IV: either a scale is generated spontaneously           
and $P_n$ falls exponentially, or else            
there appears a "singular" node with huge           
connectivity $n_{\mbox{\rm \footnotesize sing}} =           
 N (2 - \langle n \rangle_t)$. In the former case the           
graphs have a typical extension $\sim \sqrt{N}$ while in the latter case           
they are crumpled and have infinite fractal dimension. Because of the           
constant presence of the "singular" vertex the diameter of the graph           
grows slower than any power of $N$ for $N \to \infty$.           
\par           
Notice, that the condition $\beta > 2$ appears often in our discussion.           
Indeed, at least in the framework adopted           
in this paper, a scale-free graph           
with a different exponent cannot exist. Actually, the arguments used           
towards the end of Sec. IV can also be used           
 when $\langle n \rangle_t = \infty$           
and one predicts an exponentially           
falling $P_n$ at the "output".           
\par           
We have considered the stability of scale-free graphs           
with respect to possible deformations of the sequence of weights.            
Indeed, in Nature a random graph does not exist in isolation. It           
is an open system, subject to interactions with, loosely speaking, a            
heat bath. It is thus interesting to know what is expected to happen           
when this interaction becomes strong.            
\par           
Our results reflect the behavior of "typical" graphs. But very           
"atypical" values of global parameters, for example of the spectral           
dimension, can be found when a            
specific large graph is analyzed \cite{sven}.           
Our algorithm makes it possible to measure the degree of this           
"atypicality": a numerical simulation enables one to estimate the           
probability of a given fluctuation of the global parameter.           
     
\subsection{More general graph models}     
\subsubsection{Weighted connections}           
Our graph model has been constructed in such a way that the weight of           
a graph is a product of weights associated with individual nodes. A           
possible generalization could consist in replacing the starting graph           
generating function defined in (\ref{start}) by           
\begin{equation}           
Z \sim \int d^q \phi \exp{\frac{1}{\kappa} [ - \phi A \phi           
 + \sum_{n=1}^q  p_n \phi_n^n]}           
\end{equation}           
where $A$ is a positive definite matrix of rank $q$  and           
$ \phi = (\phi_1, ..., \phi_q)$. The series expansion of           
the rhs generates graphs with nodes of degree 1,2, ..., q           
and a weight $\propto A^{-1}_{mn}$ attached to every            
link connecting a pair of nodes of degree $m$ and $n$.           
The limit $q \to \infty$            
should be handled with care.           
In a study of tree graphs the limit $\kappa \to 0$            
should be taken first. The            
investigation of this model goes, however, beyond the scope of            
the present paper \cite{weighted}.           
     
\subsubsection{Matter on graphs}     
Another generalization would consist in putting "matter" fields     
on graphs. Ising spins on "small-world"     
graphs \cite{sw} were already discussed in ref. \cite{barrat}.     
Ising spins living on the nodes of quenched Feynman diagrams     
generated by $\phi^3$ or $\phi^4$ interactions     
were studied by many people (see, for example,
 \cite{des} and the long list of
references therein), in      
particular with the aim to determine     
the critical behavior of the spin system.     
It would be very interesting to extend these studies     
to scale-free graphs as defined in the present paper     
and to see, for example, what is the     
effect of changing $\beta$ and therefore also the fractal     
dimension of the graphs.      
     
\subsubsection{Graphs with cycles}     
The most important generalization consists in extending the             
study to graphs with cycles. Introduction of a            
fixed number of cycles does           
not seem to be a big deal in the present context (see, however,            
ref. \cite{jk}). For example, when $N \to \infty$ the Hausdorff     
dimension is expected to be insensitive to the presence of     
a fixed number of cycles. On the other hand,      
the case where the ratio $L/N$ is           
kept fixed at some arbitrary value when      
$N \to \infty$ is very challenging.     
\par     
There are conceptual problems related to the      
over-extensive nature      
of the ensemble. It is not difficult to count the      
number of non-degenerate      
graphs with given $N$ and $L$. It is sufficient to      
consider the number of      
distinct adjacency matrices  $C_{ij} = 1 \; \;      
\mbox{\rm or} \; \; 0$, for     
$i,j$ connected or not, respectively:     
\begin{equation}     
\mbox{\rm \# non-degenerate graphs}(N,L) =      
\left( \stackrel{\frac{1}{2}N(N-1)}{L} \right)     
\end{equation}     
In general the rhs behaves like $\exp{(\mbox{\rm const} N \log N)}$ when     
$N \to \infty$, $L/N$ fixed. A similar result is easily obtained for     
degenerate graphs. The entropy      
is not extensive. This is easy to understand.     
It is impossible to divide an arbitrary graph into      
two parts separated by      
a "boundary" of negligible measure. Any two      
nodes can interact. Hence, it is not certain,      
in general, that the bulk distribution $P_n$     
calculated in the ensemble coincides with that obtained from a single,     
sufficiently large graph. Tree graphs      
are a notable exception: in the "semiclassical limit"      
$\kappa \to 0$ the     
entropy becomes extensive.      
\par      
Our algorithm, in its most general version, generates all     
possible Feynman diagrams and can in principle be used     
to simulate arbitrary graphs. The problem is to find     
the constrains on the couplings $p_n$ insuring that the bulk     
connectivity distribution has a desired form.     
\par     
The presence of the factor $(n_k + 1) / n_j$ in     
 (\ref{metro}) means that the     
rewired nodes are sampled independently of their degree      
(without this factor      
nodes belonging to a large number of links      
would be rewired preferentially).     
If no extra check is performed before      
the Metropolis one, the algorithm does not     
know about the underlying graph structure.      
It plays with node degrees and     
nodes as it would play with balls and      
boxes in the balls-in-boxes model     
with occupation weights $p_n$. Hence a scale free $P_n$ is obtained     
setting $p_n \sim P_n$ (and 
$L=\frac{1}{2} N \langle n \rangle_P$). The graphs produced that      
way are degenerate, there are     
tadpoles and multiple connections between nodes.      
They are, in general, not connected -  it is very difficult to avoid     
producing disconnected graphs -  but     
for $\beta\leq 3$ most of nodes      
belong to the giant component.     
The conditions of its appearance are well      
understood \cite{mr,nsw}. The connectivity     
distribution within the giant component has the     
same scale-free behavior at not too small $n$     
\par      
Our present experience is     
that the algorithm has no     
difficulty to equilibrate the ensemble.     
The algorithm defines a Markov process.      
We do not have any proof that the     
next-to-largest eigenvalue of the      
transition matrix is separated from unity     
by a gap remaining finite when $N \to \infty$. But      
it seems very likely that     
it is so, since we do not expect      
any collective effects to occur.     
\par     
A different algorithm, generating graphs with     
an arbitrary connectivity, is proposed and     
used in refs. \cite{nsw,mr}. In this algorithm 
a graph is generated in two
steps. First, a list of $N$ degrees, 
$n_1, n_2, ..., n_N$, is sampled out 
of the distribution $P_n$. In a 
sense, one creates first $N$ nodes
with attached links having the other end 
free. Then, in the second step, one
connects together at random the free 
ends of links. The produced graph
is, in general, not connected and     
degenerate. The ensemble is defined by 
repeating this procedure over and over.
In the limit $N \to \infty$ the 
connectivity distribution is identical to
$P_n$ for {\em individual graphs}.
Notice, that $L=\frac{1}{2} \sum_j n_j$
is not fixed.
\par
Our approach, rooted in field theory, is different.
We start with      micro-states weighted so that the 
connectivity distribution becomes scale-free
after {\em averaging over the ensemble}.
But, as already mentioned, the scale-free 
connectivity within individual graphs is, 
strictly speaking, not guaranteed, 
except for the ensemble of trees: 
averaging within a single graph of
infinite size is, in general, not 
equivalent to averaging over the 
ensemble of graphs. This is perhaps 
a flaw, but on the whole the virtues 
of the two approaches seem complementary. 
We hope to develop this point elsewhere.     
\par     
In order to construct non-degenerate           
graphs it suffices to verify, before            
performing the Metropolis check, that            
the proposed rewiring does not produce            
tadpoles and/or multiple connections.            
We have a code doing that efficiently.     
The real problem, which we have not            
solved yet, is to find for non-degenerate graphs            
the set of couplings $p_n$            
producing the desired scale-free bulk distribution $P_n$.     
Setting $p_n=P_n$     
gives for, say, $\beta=3$ results      
that are encouraging, although the cut-off     
$n_{\mbox{\rm \footnotesize max}}$      
is smaller than in the degenerate     
case. However, when $\beta$ is      
decreased things worsen. No doubt,      
the problem deserves more theoretical and     
numerical work.             
\par           
We are           
tempted to conjecture that the degeneracy does not            
matter for large scale            
features of graphs. The absence of any natural scale suggests            
that the graphs            
are self-similar. If so, one can decimate nodes as follows: pick a node            
at random and shrink to a single point all its immediate neighbors. If            
the graph is really self-similar, decimating nodes should not alter its            
large scale features. But such a decimation will necessarily produce            
degenerate graphs. For the moment, this conjecture is a speculation.           
However, our experience with simulating random surfaces suggests that           
this possibility should not be disregarded: degenerate randomly           
triangulated surfaces - those whose duals have "tadpoles" and            
"self-energy" insertions -            
turn out to fall into           
the same universality class as the non-degenerate surfaces,           
but are much easier to simulate.           
\par     
Finally, let us note that it is straightforward to      
produce a algorithm analogous to the one proposed in this     
paper but constructing directed graphs.     
\vspace{0.4cm}           
\par           
\begin{center}           
{\bf ACKNOWLEDGMENTS}           
\end{center}           
\par           
 We  are very grateful to           
Olivier Martin for a helpful           
conversation. We would also like to thank our friends and collaborators           
Piotr Bialas, Des Johnston, Jerzy Jurkiewicz and John Wheater for           
countless discussions. This work was partially            
supported by the EC IHP network           
HPRN-CT-1999-000161, by the French-German Joint Collaboration            
Programme PROCOPE under the project 99/089 and by the           
Polish Government Project (KBN) 2P03B01917. Laboratoire de Physique           
Th\'eorique is Unit\'e Mixte du CNRS UMR 8627.

\end{multicols}         

\begin{thebibliography}{99}           
\bibitem{erdos} Erd\"os-R\'enyi theory:            
cf B. Bollob\'as, {\em Random Graphs}           
(Academic Press, London, 1985).           
\bibitem{data} A.~Broder~et~al, {\em Graph structure in the web},            
available at http://www.almaden.ibm.com/cs/k53/www9.final/           
\bibitem{simon} H.A. Simon, Biometrika {\bf 42}, 425 (1955).           
\bibitem{bar1}  A.-L. Barabasi, R. Albert, Science {\bf 286}, 509 (1999).           
\bibitem{bar2}  A.-L. Barabasi, R. Albert, H. Jeong, cond-mat/9907068.           
\bibitem{bar3}  G. Bianconi, A.-L. Barabasi, cond-mat/0011224.           
\bibitem{krap1} P.L. Krapivsky, S. Redner, F. Leyvraz, Phys. Rev. Lett.           
{\bf 85}, 4629 (2000);           
P.L. Krapivsky, S. Redner, cond-mat/0011094.           
\bibitem{doro0} S.N. Dorogovtsev, J.F.F. Mendes,            
A.N. Samukhin,  Phys. Rev. Lett. {\bf 85}, 4633 (2000).           
\bibitem{doro1} S.N. Dorogovtsev, J.F.F. Mendes,            
A.N. Samukhin, cond-mat/0011115.           
\bibitem{doro2} S.N. Dorogovtsev, J.F.F. Mendes, EuroPhys. Lett. {\bf 52},           
33 (2000);            
cond-mat/0009065; S.N. Dorogovtsev, J.F.F. Mendes, A.N. Samukhin,            
cond-mat/0009090; cond-mat/0011077.           
\bibitem{bornh} S. Bornholdt and H. Ebel, cond-mat/0008465.             
\bibitem{jan} J. Ambj\o rn, B. Durhuus, T. J\`onsson,            
Phys. Lett. {\bf B244}           
403 (1990); see also the literature on the sol-gel transition in polymer    
physics.            
\bibitem{foot1} Another possibility is to attach to each           
link a weight depending, for example, on           
the orders of the nodes it connects.           
\bibitem{foot2} Cf also refs. \cite{bar3,krap1}.           
\bibitem{bb1} P. Bialas, Z. Burda, Phys. Lett. {\bf B384}, 75 (1996).           
\bibitem{bb} P. Bialas, Z. Burda, D. Johnston,            
Nucl. Phys. {\bf B542}, 413 (1999).           
\bibitem{jk} J. Jurkiewicz, A. Krzywicki,           
 Phys. Lett. {\bf B392}, 291 (1997);            
we take this opportunity to note that the last five lines of Sec. 3.2           
are erroneous.           
\bibitem{cw} T. J\`onsson, J.F. Wheater,            
Nucl. Phys. {\bf B515}, 549 (1998);            
J.D. Correia, J.F. Wheater, Phys. Lett. {\bf B422}, 76 (1998).           
\bibitem{nsw} M.E.J. Newman, S.H. Strogatz, D.J. Watts,            
cond-mat/0007235.             
\bibitem{book} I.P. Goulden, D.M. Jackson, {\em Combinatorial Enumeration},     
John Wiley \& Sons, 1983.           
\bibitem{foot3}The recursion relation           
$$\Omega(2N,M)=\sum_{N \leq K \leq M-N} \Omega(N,K) \Omega(N,M-K)$$ is          
useful in calculating numerically, for example, the box occupation           
distribution.           
\bibitem{foot4}One readily verifies           
that the rhs of (\ref{kr}) satisfies this constraint.           
\bibitem{foot5}For integer $\beta$ the singular term is corrected by           
a logarithmic factor; hence, the critical exponents are obtained           
by taking the smooth limit $\beta \to$ integer.           
\bibitem{foot6}Actually, the observable scaling like           
$(\lambda_c - \lambda)^{-1}$ is $\langle N^p \rangle^{1/p}$ with not           
too small $p$.           
\bibitem{jan2} J. Ambj\o rn, D. Boulatov, J.L. Nielsen,            
J. Rolf, Y. Watabiki, JHEP {\bf 9802}, 010 (1998).           
\bibitem{foot7}In the code it is convenient to have           
a register of oriented links, even if we consider unoriented graphs.           
\bibitem{sven} This seems to happen in            
S. Bilke, C. Peterson, cond-mat/0103361.           
\bibitem{weighted} A related model has been considered in            
S.H. Yook, H. Jeong, A.-L. Barabasi, Y. Yu,           
cond-mat/0101309.           
\bibitem{sw} S.H. Strogatz, D.J. Watts, Nature {\bf 393}, 440 (1998).     
\bibitem{barrat} A. Barrat, M. Weigt, Eur. Phys. J. B. {13}, 547 (2000).     
\bibitem{des} D.A. Johnston, P. Plechac,  J.Phys. {\bf A30}, 7349 (1997) .     
\bibitem{mr} M. Molloy, B. Reed, Random Structures            
and Algorithms {\bf 6}, 161 (1995);           
Combinatorics, Probability and Computing {\bf 7}, 295 (1998).           
\end{thebibliography}
\end{document}